# Dementia assistive system as a dense network

Dongsoo Har

**Abstract** – As elderly population increases, portion of dementia patients becomes larger. Thus social cost of caring dementia patients has been a major concern to many nations. This article introduces a dementia assistive system operated by various sensors and devices installed in body area and activity area of patients. Since this system is served based on a network which includes a number of nodes, it requires techniques to reduce the network performance degradation caused by densely composed sensors and devices. This article introduces existing protocols for communications of sensors and devices at both low rate and high rate transmission.

*Index Terms—AAL, dementia assistive system, healthcare, dense network, MAC.*

## I. INTRODUCTION

As the population is aging, the number of chronically ill patients is enormously increasing. Particularly 10% of population aged 65 and over are suffering from dementia and Alzheimer's, which globally amount to 44 million and studies expect the number would be increased to 135milion by the 2050 [1]. Dementia refers to the deterioration of cognitive capability caused by damages to the brain cells. Dementia patients have difficulty in managing everyday life because of lack of memory and cognitive capability. Increased number of dementia patients causes enormous social cost for diagnosing, treating, and managing. Generally it is known that three people are needed to supervise one dementia patient.

The researches in various fields are progressed along with the ambient assisted living (AAL) project of EU for improving the life quality of the elderlies in 2008 [2]. Researches of early diagnosis of dementia and providing life support for dementia patients have also been conducted intensively. However, most of the researches are focused on early diagnosis or providing service with single machine. Thus there are few researches which provide complete range of dementia-related perspectives [3].

Comprehensive infrastructure, which assist patients and notify the state of patients to the remote caregivers, should be implemented to guarantee the normal life of dementia patients. To this end, events which occur inside the living space should be detected, analyzed, and notified to patients and methods to help patients deciding their own action to react various situations are required. In addition, methods to connect with the remote caregivers are required so that they can perform fast follow-up action when patients are in danger [4].

To achieve this, sensors and cameras to spot the location of patients and to recognize various situations in the living place of patients, personal monitoring equipment of patients, systems that enable patients to handle the emergency remotely, and network infrastructures that deliver the state of patients to caregivers or experts should be installed. In this system, wireless communication technology plays an important role to connect each element. This kind of system has a lot of network note and each node requires various network quality of service (QoS).

Dementia assistive system which assists normal life of dementia patient and notifies state of the patient to the remote caregiver will be explained in this article. Network characteristic of dementia assistive system and previously developed MAC protocol will be also introduced.

## II. DEMENTIA ASSISTIVE NETWORK

Dementia assistive system (DAS) is composed of assistance for patient by various devices and sensors attached on patient's body, and monitoring service which notifies the various event of patient including unusual conditions, unusual behavior, and emergency.

### A. Overview

Figure 1 represents the composition of DAS. Assistive system is composed of body area network (BAN) which is made up with sensors installed on patient's body, of activity area network(AAN) which is made up with various kinds of devices at the living place, and of tele-medical network which is designed for monitoring, remote diagnosis with medical databases and intelligent healthcare applications (pro-active side), and handling emergency (alerting side). As a system for assisted living service, DAS can be informed from devices of activity area through a device that patient has and it has a system to give orders remotely to the device. Assisted living service is functionally composed of activity area monitoring, decision supporting, and navigator and monitoring service is served through health and behavioral monitoring.  These services can be operated both separately and simultaneously.

**Activity area monitoring** – Activity area monitoring refers to a system which is designed to monitor various events in the living place of patients and help patients to perceive the events. Various kinds of sensors and cameras are installed in the activity area of patients to monitor them and the devices in the activity area analyses the state of patients and deliver the result.

**Decision supporting** - Decision supporting refers to a system which is designed to support patients in making their own decision to the events occurred in the activity area. Synchronized with the activity area monitoring, decision supporting help patients making decision about events occurred in the activity area. For example, it notifies a fire in the kitchen to patients and provides patients with various ways to deal with the emergency such as ways to extinguish fire alone or ways to call 911 to help them making decision. In addition, it provides patients with sequential procedures to achieve specific goal. For example, if electric light in specific area of the living place is broken or other devices are broken, it provides patients with a way to take an emergency measure to the device and a way to call the service center step by step. Furthermore, when guests visit the activity area, it can assist patients to do the series of procedure to check the ID of guest.

**Navigator** -Navigator is a service which is designed to help patients to navigate when they move in or outside of the active area. This service shows patients the way to their destination when they are looking for objects in the house or when they have to move for indoor or outdoor activities. In addition, when patients have to move to deal with events occurred in activity area, it shows patients the way to move along with decision supporting. Therefore, navigator service should cooperate with sensors in the activity area and location awareness system which tracks down the location of patients.

**Health & behavioral monitoring** – Elderly dementia patients have high probability of having other chronic disease as well as dementia and having an accident because of declined cognitive capability and physical ability. Therefore, DAS should check patients' heath continuously and help patients to do emergency treatment in a dangerous situation. If abnormal vital signal of patents is monitored, or if patients stay at a same place excessively long, DAS should notify the state of patients to the caregiver. Furthermore, calls for emergency treatment should be made after fall accidents are monitored which frequently happen to the old. For health and behavioral monitoring, numerous sensors installed on patients' body to collect body state information, sensors which spots patients in the activity area based on radio frequency identification (RFID)/near field communication (NFC), and cameras are used. Global navigation satellite system, sensors, and cameras are required for outdoor spotting. All these devices should be connected both by wired network and wireless network to give information to patients and their caregivers.

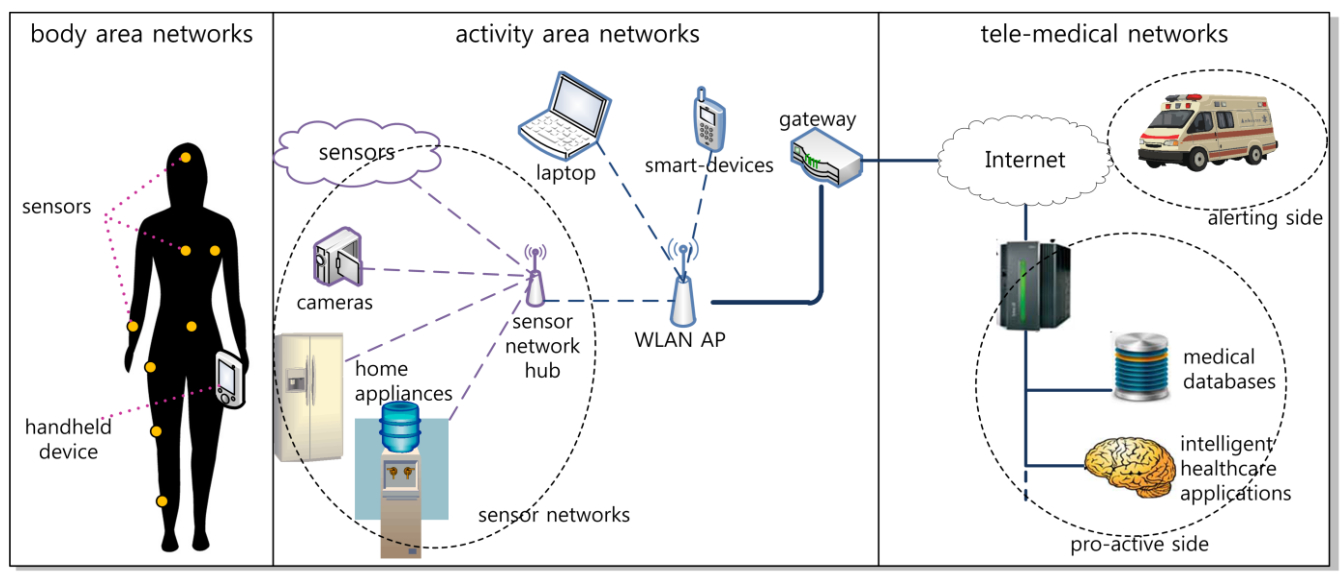

Figure 1. Dementia assistive system.

**B. Networks for dementia assistive system**

In DAS, all sorts of devices in the activity area, sensors on the body of patients, and cellphones that patients carry are linked with computer network. Wireless communication is mainly used in DAS for convenience of movement and maintenance. Each sensors and devices have various demands for wireless communication according to their own function.

BAN is aimed for collecting body state information of patients and it consists of sensors to measure heartbeat, body temperature, blood pressure, electrocardiogram (ECG), electroencephalography (EEG), and electromyography (EMG). Requirement of each sensors are described in table 1 [5]. Data rate that each sensors require is low rate lower than 1Mbps and latency has various requirement time from 100ms to 250ms or longer. Since sensors have to be used for several days or even several years without battery replacement, low power operation is also needed. As a communication protocol for low rate body area network, 802.15.4 MAC/PHY which is widely used as a protocol of wireless sensor network or 802.15.1 MAC/PHY which is developed for short-range device network can be used.

Table 1. Technical requirements of DAS sensors

| application | target data rate | latency |
|---|---|---|
| heartbeat | < 10kbps | < 250ms |
| temperature | < 10kbps | < 250ms |
| blood pressure | < 10kbps | < 250ms |
| ECG | about 72kbps | < 250ms |
| EEG | about 86.4kbps | < 250ms |
| EMG | about 1.536Mbps | < 250ms |
| audio | about 1Mbps | < 100ms |
| video | about 10Mbps | < 100ms |

Devices and sensors in the activity areas require various data rate according to their function. While sensors which measure temperature, humidity, and luminance are able to perform enough with low rate communication protocol, cameras which transmit images and sound requires higher Mbps of data rate and required Mbps is expected to be higher as image equipment develops. Furthermore hubs which collect information from sensors and cameras of BAN and ANN and transfer it to other network should provide higher rate of throughput to handle enormous traffic. Therefore it is appropriate for node, hub of network, and gateway to use high rate wireless local area network (WLAN) technique to transmit high-capacity data.

Another feature of DAS is the fact that numerous nodes exist densely in a single network. While extent of activity area in an aging society is similar to that of past, the number of node which handle wireless communication in the area is expected to explosively increase as AAL is spreading. Thus it has high probability that whole network performance fall since only nodes which occupied channel can transfer information because of the huge cost of scheduling and contention caused from high density of nodes which is the result of limited spatial coverage of network.

### C. Channel Access Protocols for DAS

In this section, features of each protocol which can be used for network of DAS will be explained

**802.15.4 Standard** - 802.15.4 which makes medium access control (MAC)/(physical)PHY layer of Zigbee is widely used to sensor network since it has high power efficiency and provide various topology. [6]. In DAS it can be used as technology which consists BAN and connect sensors in the activity area. 802.15.4 uses method which consists of time division multiple access(TDMA) and carrier sense multiple access/collision avoidance(CSMA/CA). 802.15.4 can be applied to both beacon mode and non-beacon mode. Beacon mode is a mode where sensor nodes are synchronized by beacon which is transmitted by coordinator. The area where beacon interacts with sensors is called beacon interval. This beacon interval is categorized into active portion and inactive portion and in the inactive portion use of electricity is minimized by turning off coordinator and transceiver of sensor node. Active portion, in turn, is categorized into contention access period (CAP) and contention free period (CFP). CAP and CFP consist of union of time-slot and the number of time-slot is limited up to 15. Time-slot of CFP is limited up to 7. In CAP, each node transmits information by CSMA/CA method where nodes access to channel by carrier sensing. In CFP, where competitive transmit is made, only a node which is appointed by coordinator transmit information.

In non-beacon mode, each node uses CSMA/CA for transmission and there is no distinction between active portion and inactive portion since there is no concept of beacon interval.

While 802.15.4 has high electricity efficiency, since the number of slot of CFP is limited up to 7 most sensors have to compete for channel access in CAP. This, in turn, ruins the QoS and fairness among sensors.

**802.15.1 standard** - 802.15.1 which composes MAC/PHY layer of Bluetooth can provide up to 3Mbps data rate so it can transmit high-capacity information compared to 802.15.4 [7]. It uses polling method for channel access method. Polling method is the system where nodes send request to coordinator before start transmit and start transmit after receiving acknowledge from coordinator. In addition, it makes frequency hopping to 79 channels 1600 times per 1 second to restrain interference from other networks around. When adaptive frequency hopping is applied, if collision is detected at the channel which was hopped, the channel is blocked to prevent collision.

However, it has its weakness at the fact that only seven sensors can be linked to one master. In addition frequency hopping between 79 channels makes total 79MHz band. This, in turn, may cause interference to the other networks which use frequency of 2.4GHz. Furthermore, if a communication system which has broad channel band (for example, 802.11 standards which has maximum channel bandwidth of 40MHz at 2.4GHz) is being used around, if frequency hopping is committed, many channels can be unserviceable because of collision.

**802.11 standard -** There exist various standard of MAC/PHY such as 802.11b/g/n/ac/ad and MAC/PHY is used at Wi-Fi [8-10]. Channel access is based on CSMA/CA and it can provide both priority based access and contention-free access optionally. Huge decline of efficiency of frame occurs when contention-based access is implemented since it uses binary exponential backoff mechanism to avoid collision. Especially average throughput of nodes in network declines rapidly as the number of nodes increases. For example, in case of 802.11n which is most commonly used, if 10 nodes exist in the channel which has the capacity of 72Mbps, average throughput declines 3.5Mbps per one node [11]. In addition, it consumes more electricity than 802.15.4, and 802.15.1 and because of its comparably huge output; it has high probability of causing interference to other communication techniques which use same frequency.

Therefore, 802.15.4 is appropriate for DAS because it can connect numerous client nodes to network and can operate with low power in the low rate communication such as BAN and AAN. 802.11 standard is appropriate for transmission of video data, and gateway communication which connect BAN and AAN to other network. In this case, BAN and ANN are recommended to use frequency of 2.4GHz and 802.11 is recommended to use frequency of 5GHz which n or ac provides to solve to interference problem of BAN, AAN, and WLAN.

## V. CONCLUSIONS

This article introduced the dementia assistive system to assist the ordinary life of dementia patient and to monitor the abnormal state of patients. Characteristics of DAS network composed of many nodes are briefly addressed. Data transmission rate of the sensors and devices can be ranging from a few kbps to tens of Mbps. It is also noted that much higher data rate system such as OFDM system[12-13] is also applicable when it is required. Representative channel access protocols for low rate and high rate are shortly discussed.